\documentclass[11pt]{article}
\usepackage{amssymb}
\usepackage{citesort}
\usepackage{graphics,graphicx}

\def\ket#1{\left|#1\right\rangle}

\begin{document}
\begin{titlepage}
\thispagestyle{empty}

\begin{flushright}
                arXiv:0711.1618
\end{flushright}
\vskip 2cm

\begin{center}

{\LARGE\bf\sf Conformal blocks related to the R-R states in the $\hat c =1$ SCFT} \\

\end{center}
\vskip 1cm

\begin{center}

    {\large\bf\sf
    Leszek Hadasz${}^\dag$\footnote{\emph{e-mail}: hadasz@th.if.uj.edu.pl}$\!\!\!\!,\ \,$
    Zbigniew Jask\'{o}lski${}^\ddag$\footnote{\emph{e-mail}: jask@ift.uni.wroc.pl}
    and
    Paulina Suchanek${}^\dag$\footnote{\emph{e-mail}: suchanek@th.if.uj.edu.pl}
    }
     \\
\vskip 3mm
    ${}^\dag$ M. Smoluchowski Institute of Physics,
    Jagiellonian University, \\
    Reymonta 4,
    30-059~Krak\'ow, Poland, \\

\vskip 3mm
    ${}^\ddag$ Institute of Theoretical Physics,
    University of Wroc{\l}aw, \\
    pl. M. Borna, 50-204~Wroc{\l}aw, Poland. \\
\end{center}

\vskip 1cm

\begin{abstract}
We derive an explicit form of a family of four-point Neveu-Schwarz
blocks with $\hat c =1,$ external weights $\Delta_i = \frac18$ and
arbitrary intermediate weight $\Delta.$ The derivation is based on a
set of identities obeyed in the free superscalar theory by
correlation functions of fields satisfying Ramond condition with
respect to the  bosonic (dimension 1) and the fermionic (dimension
$\frac12$) currents.
\end{abstract}

\vspace{\fill}


\vspace*{5mm}
\end{titlepage}
\newpage
\vskip 1cm
\section{Introduction}
Conformal field theory proved to be very efficient in describing
second order phase transitions in two-dimensional system and is
accepted as a language of string theory. Correlation functions in
CFT can be expressed as sums (or integrals) of three-point
coupling constants and the conformal blocks, fully determined by
the symmetry alone. The basic role played by the blocks has been
recognized since the appearance of the ground-breaking BPZ work
\cite{Belavin:1984vu}.

In spite of the progress achieved over the years the analytic form
of the general block is unknown and the explicit examples are
mostly limited to the blocks corresponding to degenerate
representations of the underlying Virasoro algebra or the blocks
appearing in the correlation functions of free fields. An
interesting example of the latter kind is
a family of conformal blocks related to the Ramond states of a
free scalar fields \cite{Zamolodchikov:1,Zamolodchikov:A}. The
exact analytic results of \cite{Zamolodchikov:1,Zamolodchikov:A}
play an essential r\^ole  in developing the so called elliptic
recursion representation of the general conformal block
\cite{Zamolodchikov:3,Zamolodchikov:4}.

Recently the recursion representations have been also worked out for
the super-conformal blocks related to the Neveu-Schwarz algebra
\cite{Hadasz:2006sb,Belavin:2006,Belavin:2007}. In particular the
elliptic recursion has been conjectured and
 used   for the numerical
verification of the consistency of $N =1$ super-Liouville theory
\cite{Belavin:2007}. One of the missing steps of a possible
rigorous proof of this conjecture is a derivation of the large
intermediate weight asymptotic of the general NS superconformal
block. It can be obtained by a method parallel to the one used in
\cite{Zamolodchikov:3} in the Virasoro case. This however requires
explicit analytic formulae
 of certain superconformal blocks and was the main motivation
behind the present work.

The organization of the paper is as follows. In Section 2 we
briefly describe the structure of $\hat c=1$ free superscalar
theory extended by the Ramond states both in the bosonic and the
fermionic sector. In Section 3 we derive a set of relations for
the 4-point functions which are used in Section 4 to derive a
closed system of 6  equations for superconformal blocks. The
solutions to these equations provide new nontrivial examples of NS
superconformal blocks and are the main result of the present work.

\section{Holomorphic currents}
The free bosonic current  $j(z)$ (with conformal weights $\Delta = 1,\, \bar\Delta = 0$)
satisfies the relation:
$$
j(z)j(z') \sim {1\over (z-z')^2}.
$$
Following \cite{Zamolodchikov:A} one may consider two types of states
of a free scalar field:
the NS states $\ket{\xi}_{NS}$ for which
\begin{equation}
\label{CCRNS}
j(z)\ket{\xi}_{NS}
=\sum\limits_{n\in \mathbb{Z}} z^{-n-1}j_n
\ket{\xi}_{NS},\;\;\;\;\;\;\;
[j_n,j_m] = m \delta_{n+m},
\end{equation}
and the R states $\ket{\xi}_{R}$ characterized by
\begin{equation}
\label{CCRR}
j(z)\ket{\xi}_{R}
=\sum\limits_{k\in \mathbb{Z}+{1\over 2}} z^{-k-1}j_k
\ket{\xi}_{R},\;\;\;\;\;\;\;
[j_k,j_l] = k \delta_{k+l}.
\end{equation}
The space of states $\cal B$
is (by construction)
a direct sum
$$
{\cal B} =\left( \bigoplus\limits_p {\cal B}^{NS}_p \right)\oplus {\cal B}^{R}
$$
where
 ${\cal B}^{NS}_p$ are the NS  current modules
defined as a highest weight representations of the algebra
(\ref{CCRNS})
with the highest weight state
\begin{equation}
\label{groud:j:NS}
\begin{array}{llllllllllllllllll}
j_0\ket{p}_{NS} = p\ket{p}_{NS},\;&&\;j_n\ket{p}_{NS} = 0,\;&&\;n\in \mathbb{N},
\end{array}
\end{equation}
and
 ${\cal B}^{R}$ is the R current module defined as a highest weight representation of the algebra
 (\ref{CCRR})
with the highest weight state
\begin{equation}
\label{groud:j:R}
\begin{array}{llllllllllllllllll}
&&j_k\ket{0}_{R} = 0,\;&&\;k\in \mathbb{N}-{1\over 2}.
\end{array}
\end{equation}
We shall use a similar construction for the free fermion current, defined by the OPE
$$
\psi(z)\psi(z')\sim {1\over z-z'}\ .
$$
In this case
\begin{eqnarray}
\label{CARNS}
\psi(z)\ket{\zeta}_{NS}
&=&\sum\limits_{k\in \mathbb{Z}+{1\over 2}} z^{-k-{1\over 2}}\psi_k
\ket{\zeta}_{NS},\;\;\;\;\;\;\;
\{\psi_k,\psi_l\} =  \delta_{k+l},
\\
\label{CARR}
\psi(z)\ket{\zeta}_{R}
&=&\sum\limits_{n\in \mathbb{Z}} z^{-n-{1\over 2}}\psi_n
\ket{\zeta}_{R},\;\;\;\;\;\;\;
\{\psi_n,\psi_m\} =  \delta_{n+m}.
\end{eqnarray}
The space of states $\cal F$
is a direct sum of the  fermionic NS current module  ${\cal F}^{NS}$ and the fermionic R current module
 ${\cal F}^R$ built on the
highest weight states $\ket{0}_{NS}$ and $\ket{+}_{R}$ of the algebras (\ref{CARNS}) and (\ref{CARR}), respectively,
defined by the relations:
$$
\begin{array}{llllllllllllllll}
&&&&\psi_k\ket{0}_{NS} &=& 0,\;&&k\in \mathbb{N}-{1\over 2},\\
\psi_0\ket{+}_{R}&=&{1\over \sqrt{2}}\ket{-}_{R},&&\psi_n\ket{+}_{R} &=& 0,\;&&n\in \mathbb{N}.
\end{array}
$$

The tensor product ${\cal B}\otimes {\cal F}$ decomposes into the direct sum
$$
{\cal B}\otimes {\cal F}
=
\left[\left( \bigoplus\limits_p {\cal B}^{NS}_p \otimes
{\cal F}^{NS} \right)\oplus {\cal B}^R\otimes {\cal F}^R\right]
\oplus
\left[\left( \bigoplus\limits_p {\cal B}^{NS}_p \otimes
{\cal F}^R \right)\oplus {\cal B}^R\otimes {\cal F}^{NS}\right]
$$
of highest weight supercurrent modules.
The Sugawara construction
\begin{eqnarray*}
T(z)&=& {1\over 2} :\!j(z)j(z)\!: - {1\over 2} :\!\psi(z)\partial \psi(z)\!:\, ,\\
S(z)&=& j(z)\psi(z),
\end{eqnarray*}
defines on the first summand a free field representation of the NS superconformal algebra
with the central charge $\hat c = {2\over 3} c = 1$.
In this sector
$$
T(z) = \sum\limits_{n\in \mathbb{Z}}z^{-n-2}L_n,\;\;\;
\;\;\;
S(z)= \sum\limits_{k\in \mathbb{Z}+{1\over 2}}z^{-k-{3\over 2}}S_k,
$$
where
\begin{eqnarray*}
L_0
&=&
{1\over 2}j_0^2 + \sum\limits_{n\in \mathbb{N}}j_{-n}j_n +
\sum\limits_{k\in \mathbb{N}-{1\over 2}}(k+\textstyle{1\over 2})\psi_{-k}\psi_k,
\\
L_m
&=&
{1\over 2}\sum\limits_{n\in \mathbb{Z}}j_{m-n}j_n
+{1\over 4}\sum\limits_{k\in \mathbb{Z}+{1\over 2}}(2k-m)\psi_{m-k}\psi_k,
\hskip 10mm m \neq 0,
\\
S_k
&=&
\sum\limits_{n\in \mathbb{Z}}j_{n}\psi_{k-n},
\end{eqnarray*}
on the subspace $ \bigoplus\limits_p {\cal B}^{NS}_p \otimes
{\cal F}^{NS}$ and
\begin{eqnarray}
\label{LS:RR}
\nonumber
L_0
&=&
\sum\limits_{k\in \mathbb{N}-{1\over 2}}j_{-k}j_k +
\sum\limits_{n\in \mathbb{N}}(n+{1\over 2})\psi_{-n}\psi_n +{1\over 8},
\\
L_m
&=&
{1\over 2}\sum\limits_{k\in \mathbb{Z}+\textstyle{1\over 2}}j_{m-k}j_k\
+{1\over 4}\sum\limits_{n\in \mathbb{Z}}(2n-m)\psi_{m-n}\psi_n,
\hskip 1cm
m\neq 0,
\\
\nonumber
S_k
&=&
\sum\limits_{n\in \mathbb{Z}}\psi_{n}j_{k-n}
\end{eqnarray}
on ${\cal B}^{R}\otimes {\cal F}^{R}.$

One easily verifies  that the  NS-NS supercurrent  module ${\cal B}^{NS}_p \otimes
{\cal F}^{NS}$
is an NS superconformal Verma module with the conformal weight
$
\Delta_p = {p^2\over 2}.
$
We shall denote the corresponding superprimary field by $\varphi_p(z)$:
$$
\varphi_p(0)\,\ket{0} = \ket{p}_{NS}\otimes\ket{0}_{NS}\equiv \nu_p,
$$
where
$\ket{0}=\ket{0}_{NS}\otimes\ket{0}_{NS}\in {\cal B}_0^{NS}\otimes {\cal F}^{NS} $
is the ``true'' vacuum.

On the other hand in the R-R supercurrent   module ${\cal B}^{R}\otimes {\cal F}^{R} $
one has two superprimary states at each ${n(n+1)\over 2}$ level. Indeed,
since all $\hat c =1$,  $\Delta_n ={1\over 2}\left(n+{1\over 2}\right)^2$ NS superconformal
 Verma modules are not degenerate  the superconformal content of the R-R module
can be inferred from the ratio
$$
{\chi_{RR} (t)
\over
\chi_c (t)}
\; = \;
2\sum\limits_{n=0}^\infty t^{n(n+1)\over 2}
$$
where $\chi_{RR}(t) $ is  the character of ${\cal B}^{R}\otimes {\cal F}^{R},$
$$
\chi_{RR} (t) = 2 t^{1\over 8}\prod\limits_{k=1}^\infty {1+t^k\over 1- t^{2k-1\over 2}},
$$
and
$$
\chi_c (t) = t^{1\over 8}\prod\limits_{k=1}^\infty {1+ t^{2k-1\over 2}\over 1-t^k }
$$
is the character of the superconformal NS module.

The R-R module is thus  a direct sum of irreducible NS superconformal Verma modules with conformal
weights
$$
\Delta_n ={1\over 2}\left(n+{1\over 2}\right)^2,\;\;\;\;\;\;n=0,1,\dots,
$$
each weight appearing twice in the sum. We shall denote the corresponding superprimary fields by
$
\chi^\pm_n(z) .
$
In particular
$$
\chi^\pm_0(0)\, \ket{0} = \ket{0}_{R}\otimes\ket{\pm}_{R}\equiv \chi_0^\pm.
$$
It is also easy to check that the super-primary field with the (left) weight $\Delta_1$ can
be expressed as:
\begin{equation}
\label{chi:1}
\chi^\pm_1(z) \;\ = \;\ \frac12\left(j_{-\frac12}^2 - \psi_{-1}\psi_0\right)\chi_{0}^{\pm}(z).
\end{equation}

All the notions related to the superconformal algebra like primary and descended fields,
families, blocks,
operators-states correspondence etc, have their counterparts in  the case of the supercurrent algebra.
One can show in particular that the 3-point function of  fields from arbitrary supercurrent
families factorizes into a product of 3-point supercurrent blocks:
\begin{eqnarray}
\label{eta:def}
&&\hspace{-100pt}
\langle
\,
\phi_3(\xi_3,\bar\xi_3 |z_3,\bar z_3)
\phi_2(\xi_2,\bar\xi_2 |z_2,\bar z_2)
\phi_1(\xi_1,\bar\xi_1 |z_1,\bar z_1)
\,\rangle \;=\\ \nonumber
&&
\hskip 50pt = \;
\eta{_{z_3}}{_{z_2}}{_{z_1}} (\xi_3,\xi_2,\xi_1)\,
\eta{_{\bar z_3}}{_{\bar z_2}}{_{\bar z_1}}
(\bar \xi_3,\bar \xi_2,\bar \xi_1).
\end{eqnarray}
The form $\eta$ is a nontrivial extension of the 3-point superconformal block
in the case of one arbitrary NS-NS and two R-R supercurrent modules:
$$
\eta{_{z_3}}{_{z_2}}{_{z_1}} (\xi,\zeta,\zeta'),\;\;\;\;\;\;
\xi \in {\cal B}_p^{NS}\otimes {\cal F}^{NS},\;\;\;\;\;\;\zeta,\zeta'
\in {\cal B}^{R}\otimes {\cal F}^{R}.
$$
It is uniquely determined by Ward identities for currents $j(z)$ and $\psi(z)$.
Since in the free superscalar theory the left and the right fermionic parities\footnote{In all supercurrent
modules the parity is defined by the number of fermionic excitations.}
are independently preserved the form $\eta$ is necessarily  even, i.e.\
it vanishes identically if total parity of all arguments is odd.

If the states $\zeta,\zeta'$ belong to  definite superconformal Verma modules the form $\eta$ can be also
calculated using the superconformal Ward identities. For instance, for even vectors
$$
\nu_{p,{\scriptscriptstyle KM}}
\; = \;
S_{-K} L_{-M}\nu_{p}
\; \equiv \;
S_{-k_i}\ldots S_{-k_1}L_{-m_j}\ldots L_{-m_1}\nu_{p}\,,\;\;\;\;|K|\in \mathbb{N}\cup\{0\},
$$
one has:
\begin{eqnarray}
\label{factorization:even}
\nonumber
\eta{_{z_3}}{_{z_2}}{_{z_1}}
(\nu_{p,{\scriptscriptstyle KM}},\chi^{\pm}_m,\chi^{\pm}_n)
& = &
\eta_{\infty\: 1\: 0}
(\nu_{p},\chi^{\pm}_m,\chi^{\pm}_n)\,
\rho^{\Delta_p\,\Delta_m\,\Delta_n}_{\,z_3\;\ z_2\;\ z_1}
(\nu_{p,{\scriptscriptstyle KM}},\chi_m,\chi_n),
\\[-5pt]
\\[-5pt]
\nonumber
\eta{_{z_3}}{_{z_2}}{_{z_1}}
(\nu_{p,{\scriptscriptstyle KM}},S_{-\frac12}\chi_m^{\mp},\chi_n^{\pm})
& = &
\eta_{\infty\: 1\: 0}
(\nu_{p},S_{-\frac12}\chi_m^{\mp},\chi_n^{\pm})\,
\rho^{\Delta_p\,\Delta_m\,\Delta_n}_{\,z_3\;\ z_2\;\ z_1}
(\nu_{p,{\scriptscriptstyle KM}},*\chi_m,\chi_n),
\end{eqnarray}
and for odd ones ($|K|\in \mathbb{N}-{1\over 2}$):
\begin{eqnarray}
\label{factorization:odd}
\nonumber
\eta{_{z_3}}{_{z_2}}{_{z_1}}
(\nu_{p,{\scriptscriptstyle KM}},\chi^{\mp}_m,\chi^{\pm}_n)
& = &
\eta_{\infty\: 1\: 0}
(\nu_{p},S_{-\frac12}\chi^{\mp}_m,\chi^{\pm}_n)\,
\rho^{\Delta_p\,\Delta_m\,\Delta_n}_{\,z_3\;\ z_2\;\ z_1}
(\nu_{p,{\scriptscriptstyle KM}},\chi_m,\chi_n),
\\[-5pt]
\\[-5pt]
\nonumber
\eta{_{z_3}}{_{z_2}}{_{z_1}}
(\nu_{p,{\scriptscriptstyle KM}},S_{-\frac12}\chi_m^{\pm},\chi_n^{\pm})
& = &
\eta_{\infty\: 1\: 0}
(\nu_{p},\chi_m^{\pm},\chi_n^{\pm})\,
\rho^{\Delta_p\,\Delta_m\,\Delta_n}_{\,z_3\;\ z_2\;\ z_1}
(\nu_{p,{\scriptscriptstyle KM}},*\chi_m,\chi_n).
\end{eqnarray}
The form $\rho$ in the formulae above is the normalized 3-point superconformal block
introduced in \cite{Hadasz:2006sb} and $\chi_m$ stands for the highest weight state
in the superconformal Verma module with the central charge $c ={3\over 2}$ and
the conformal weight $\Delta_m = \frac12\left(m+\frac12\right)^2$.

\section{Relations for the correlation functions of R-R fields }

We derive now equations for some 4-point correlation functions which will be
used in the next section to obtain
equations for the conformal blocks
${\cal F}^{1}_{\Delta_p}\!
\left[^{\underline{\hspace{3pt}}\,\Delta_3 \;\underline{\hspace{3pt}}\,\Delta_2}_{\hspace{3pt}\,\Delta_4\;\hspace{3pt}\, \Delta_1} \right]\!(z)$
and
${\cal F}^{\frac12}_{\Delta_p}\!
\left[^{\underline{\hspace{3pt}}\,\Delta_3 \;\underline{\hspace{3pt}}\,\Delta_2}_{\hspace{3pt}\,\Delta_4\;\hspace{3pt}\, \Delta_1} \right]\!(z).$
The derivation is based on an supersymmetric
extension of the  technique \cite{Zamolodchikov:A}.

Consider the correlation function with an arbitrary pattern of upper signs
\begin{eqnarray}
\label{four:chi:and:j0}
\Big\langle\chi^{\pm}_0(z_4)\chi^{\pm}_0(z_3)\,j_0\,\chi^{\pm}_0(z_2)\chi^{\pm}_0(z_1)\Big\rangle
\hskip -2pt
& \stackrel{\rm def}{=} &
\hskip -10pt
\oint\limits_{{\cal C}_{[z_2,z_1]}}
\hskip -10pt
\frac{d\xi}{2\pi i}\
\Big\langle j(\xi)\chi^{\pm}_0(z_4)\chi^{\pm}_0(z_3)\chi^{\pm}_0(z_2)\chi^{\pm}_0(z_1)\Big\rangle,
\end{eqnarray}
where the positively oriented integration contour encloses points $z_1$ and $z_2.$ Equations
(\ref{CCRR}) and (\ref{groud:j:R}) give the OPE of the primary field $\chi^\pm_0(z)$ and
the current $j(\xi)$:
\begin{equation}
\label{OPE1}
j(\xi)\chi^\pm_0(z) \;\ \sim \;\ \frac{1}{\sqrt{\xi -z}}\, j_{-\frac12}\chi^\pm_0(z).
\end{equation}
The function
\[
\Big\langle j(\xi)\chi^{\pm}_0(z_4)\chi^{\pm}_0(z_3)\chi^{\pm}_0(z_2)\chi^{\pm}_0(z_1)\Big\rangle\,
\sqrt{(\xi-z_4)(\xi-z_3)(\xi-z_2)(\xi-z_1)}
\]
is therefore a single valued, holomorphic function of $\xi.$ Since any correlator
of $j(\xi)$ with no operator insertion at infinity falls like $\xi^{-2}$ for large $\xi$ this function
is a constant, hence
\[
\Big\langle j(\xi)\chi^{\pm}_0(z_4)\chi^{\pm}_0(z_3)\chi^{\pm}_0(z_2)\chi^{\pm}_0(z_1)\Big\rangle
\;\ = \;\
\frac{A(z_i)}{\sqrt{(\xi-z_4)(\xi-z_3)(\xi-z_2)(\xi-z_1)}}.
\]
Expanding the r.h.s. of this equation around $\xi = z_2$ and comparing the result with the OPE
(\ref{OPE1}) we get
\[
A(z_i)
\;\ = \;\
\sqrt{z_{21}z_{23}z_{24}}\,
\Big\langle \chi^{\pm}_0(z_4)\chi^{\pm}_0(z_3)j_{-\frac12}\chi^{\pm}_0(z_2)\chi^{\pm}_0(z_1)\Big\rangle.
\]
The integral on the r.h.s. of (\ref{four:chi:and:j0}) can be now performed explicitly and
(\ref{four:chi:and:j0}) takes the form:
\begin{equation}
\label{first}
\Big\langle\chi^{\pm}_0(z_4)\chi^{\pm}_0(z_3)\,j_0\,\chi^{\pm}_0(z_2)\chi^{\pm}_0(z_1)\Big\rangle
\; = \; \sqrt{z_{21}z_{23}z_{24}}\,
{\cal K}(z_i)\,
\Big\langle \chi^{\pm}_0(z_4)\chi^{\pm}_0(z_3)j_{-\frac12}\chi^{\pm}_0(z_2)\chi^{\pm}_0(z_1)\Big\rangle,
\end{equation}
where
\[
z \;\ = \;\ \frac{z_{21}z_{43}}{z_{31}z_{42}}
\]
is the four-point projective invariant and
\[
{\cal K}(z_i)
\;\ =
\oint\limits_{{\cal C}_{[z_2,z_1]}}
\hskip -10pt
\frac{d\xi}{2\pi i}\
\frac{1}{\sqrt{(\xi-z_1)(\xi-z_2)(\xi-z_3)(\xi-z_4)}}
\;\ = \;\
\frac{2 K(z)}{\pi\sqrt{z_{31}z_{42}}}\,,
\]
with
\[
K(z) \;\ = \;\ \int\limits_{0}^{1}\! \frac{dt}{\sqrt{\left(1-t^2\right)\left(1-t^2z\right)}}
\]
being the complete elliptic integral of the first kind.

Using the algebra of modes $j_k$ and $\psi_m$ and relations (\ref{LS:RR}) and (\ref{chi:1}) one gets:
\begin{eqnarray}
\label{OPE2:a}
j(\xi)\,j_{-\frac12}\chi_0^{\pm}(z)
& \sim &
\frac{1}{2(\xi-z)^\frac32}\,\chi_0^{\pm}(z)
+
\frac{1}{\sqrt{\xi-z}}\,L_{-1}\chi_0^{\pm}(z)
+
\frac{1}{\sqrt{\xi-z}}\,\chi^{\pm}_1(z).
\end{eqnarray}
It then follows from OPE-s (\ref{OPE1}) and (\ref{OPE2:a}) that
\[
\Big\langle j(\xi)\chi^{\pm}_0(z_4)\chi^{\pm}_0(z_3)j_{-\frac12}\chi^{\pm}_0(z_2)\chi^{\pm}_0(z_1)\Big\rangle\,
\sqrt{(\xi-z_4)(\xi-z_3)(\xi-z_2)(\xi-z_1)}\,,
\]
considered as a function of $\xi,$ is  holomorphic  on ${\mathbb C}\setminus \{z_2\},$ vanishes at
infinity like $\xi^{-2}$ and has a simple pole at $\xi = z_2,$
hence
\begin{eqnarray}
\label{second:temp}
&&
\hskip -3cm
\Big\langle j(\xi)\chi^{\pm}_0(z_4)\chi^{\pm}_0(z_3)j_{-\frac12}\chi^{\pm}_0(z_2)\chi^{\pm}_0(z_1)\Big\rangle
\; =
\\
\nonumber
& = &
\frac{1}{\sqrt{(\xi-z_4)(\xi-z_3)(\xi-z_2)(\xi-z_1)}}
\left(\frac{B(z_i)}{\xi - z_2} + C(z_i)\right).
\end{eqnarray}
Expanding the r.h.s. of (\ref{second:temp}) around $\xi = z_2$ and comparing with (\ref{OPE2:a}) we get:
\begin{eqnarray*}
B(z_i)
& = &
\frac12\sqrt{z_{21}z_{23}z_{24}}\,
\Big\langle
\chi_0^{\pm}(z_4)\chi_0^{\pm}(z_3)\chi_0^{\pm}(z_2)\chi_0^{\pm}(z_1)
\Big\rangle,
\\[10pt]
C(z_i)
& = &
\sqrt{z_{21}z_{23}z_{24}}
\left[
\Big\langle
\chi_0^{\pm}(z_4)\chi_0^{\pm}(z_3)L_{-1}\chi_0^{\pm}(z_2)\chi_0^{\pm}(z_1)
\Big\rangle
+
\Big\langle
\chi_0^{\pm}(z_4)\chi_0^{\pm}(z_3)\chi_1^{\pm}(z_2)\chi_0^{\pm}(z_1)
\Big\rangle
\right.
\\[10pt]
&&
\hskip 3.7cm
\left.
+\;\
\frac14
\left(
\frac{1}{z_{21}} + \frac{1}{z_{23}} + \frac{1}{z_{24}}
\right)
\Big\langle
\chi_0^{\pm}(z_4)\chi_0^{\pm}(z_3)\chi_0^{\pm}(z_2)\chi_0^{\pm}(z_1)
\Big\rangle
\right].
\end{eqnarray*}
Inserting this into (\ref{second:temp}) and integrating along ${\cal C}_{[z_2,z_1]}$
one obtains:
\begin{eqnarray}
\label{second}
&&
\nonumber
\hskip -2cm
\Big\langle
\chi_0^{\pm}(z_4)\chi_0^{\pm}(z_3)\,j_0\,j_{-\frac12}\chi_0^{\pm}(z_2)\chi_0^{\pm}(z_1)
\Big\rangle
\\[10pt]
& = &
\Big(z_{21}z_{32}z_{42}\Big)^{\frac14}
\frac{\partial}{\partial z_2}
\left[
\Big(z_{21}z_{32}z_{42}\Big)^{\frac14}\,
{\cal K}(z_i)\,
\Big\langle
\chi_0^{\pm}(z_4)\chi_0^{\pm}(z_3)\chi_0^{\pm}(z_2)\chi_0^{\pm}(z_1)
\Big\rangle
\right]
\\[10pt]
\nonumber
& + &
\Big(z_{21}z_{32}z_{42}\Big)^{\frac12}\,
{\cal K}(z_i)\,
\Big\langle
\chi_0^{\pm}(z_4)\chi_0^{\pm}(z_3)\chi_1^{\pm}(z_2)\chi_0^{\pm}(z_1)
\Big\rangle
\end{eqnarray}
where we  used the CWI
\[
L_{-1}\chi_0^{\pm}(z) = \partial_z\chi_0^{\pm}(z).
\]

Another set of equations for correlation functions can be obtained using the
OPE-s:
\begin{eqnarray}
\label{OPE2:b}
\sqrt{2}\psi(\xi)\,\chi_0^{\pm}(z)
& \sim &
 \frac{1}{\sqrt{\xi -z}}\, \chi^\mp_0(z)
,
\\[10pt]
\label{OPE2:c}
\sqrt{2}\psi(\xi)\,\chi_1^{\pm}(z)
& \sim &
-\frac{1}{2(\xi-z)^{\frac32}}\,\chi_0^{\mp}(z)
+
\frac{1}{\sqrt{\xi-z}}\,L_{-1}\chi_0^{\mp}(z),
\end{eqnarray}
which can be easily derived from
 the algebra of modes $\psi_m$ together with the relations (\ref{LS:RR}) and (\ref{chi:1}).
It follows from (\ref{OPE2:b}) and (\ref{OPE2:c}) that
\begin{equation}
\label{collerator}
\frac{\left\langle
\sqrt2\psi(\xi)\,
\chi_0^{\pm}(z_4)\chi_0^{\pm}(z_3)\chi_m^{\pm}(z_2)\chi_0^{\pm}(z_1)
\right\rangle}{\sqrt{(\xi-z_1)(\xi-z_2)(\xi-z_3)(\xi-z_4)}}
\end{equation}
is an  analytic function of $\xi,$  with poles at the locations $z_i$
and vanishing at infinity faster than $\xi^{-1}$.
Sum of its
residues must therefore vanish and with the help of
(\ref{OPE2:b}) we get in particular
\begin{eqnarray}
\label{third_a}
\nonumber
0&=&
\nonumber
{
\left\langle
\chi_0^{-}(z_4)\chi_0^{+}(z_3)\chi_0^{+}(z_2)\chi_0^{-}(z_1)
\right\rangle
\over
\sqrt{z_{41}z_{42}z_{43}}}
+
{
\left\langle
\chi_0^{+}(z_4)\chi_0^{-}(z_3)\chi_0^{+}(z_2)\chi_0^{-}(z_1)
\right\rangle
\over
\sqrt{z_{31}z_{32}z_{34}}}
\\[10pt]
&+&
{
\left\langle
\chi_0^{+}(z_4)\chi_0^{+}(z_3)\chi_0^{-}(z_2)\chi_0^{-}(z_1)
\right\rangle
\over
\sqrt{z_{21}z_{23}z_{24}}}
+
{
\left\langle
\chi_0^{+}(z_4)\chi_0^{+}(z_3)\chi_0^{+}(z_2)\chi_0^{+}(z_1)
\right\rangle
\over
\sqrt{z_{12}z_{13}z_{14}}}
\end{eqnarray}
for $m=0$ and
\newpage
\begin{eqnarray}
\label{third_b}
\nonumber
&&\hspace{-100pt}
-\left(z_{21}z_{23}z_{24}\right)^{-{3\over 4}}
{\partial \over \partial z_2} \left[
(z_{21}z_{23}z_{24})^{{1\over 4}}
\left\langle
\chi_0^{+}(z_4)\chi_0^{+}(z_3)\chi_1^{-}(z_2)\chi_0^{-}(z_1)
\right\rangle
\right]\;=
\\
&=&\left(z_{41}z_{42}z_{43}\right)^{-{1\over 2}}
\left\langle
\chi_0^{-}(z_4)\chi_0^{+}(z_3)\chi_1^{+}(z_2)\chi_0^{-}(z_1)
\right\rangle
\\
\nonumber
&+&\left(z_{31}z_{32}z_{34}\right)^{-{1\over 2}}
\left\langle
\chi_0^{+}(z_4)\chi_0^{-}(z_3)\chi_1^{+}(z_2)\chi_0^{-}(z_1)
\right\rangle
\\
\nonumber
&+&\left(z_{12}z_{13}z_{14}\right)^{-{1\over 2}}
\left\langle
\chi_0^{+}(z_4)\chi_0^{+}(z_3)\chi_1^{+}(z_2)\chi_0^{+}(z_1)
\right\rangle
\end{eqnarray}
for $m=1$.

One more set of equations can be derived integrating (\ref{second:temp}) around $z_3$:
\begin{eqnarray}
\label{czwarty}
&& \hspace{-40pt}
\left\langle
\chi_0^{\pm}(z_4)\,j_{-{1\over 2}}\chi_0^{\pm}(z_3)\,j_{-{1\over 2}}\chi_0^{\pm}(z_2)\chi_0^{\pm}(z_1)
\right\rangle \\
&=& \nonumber
\oint\limits_{z_3}
\frac{d\xi}{2\pi i}\ \frac{1}{\sqrt{\xi - z_3}}
\left\langle
j(\xi)\, \chi_0^{\pm}(z_4)\chi_0^{\pm}(z_3)\,j_{-{1\over 2}}\chi_0^{\pm}(z_2)\chi_0^{\pm}(z_1)
\right\rangle
\\ \nonumber
&=& \sqrt{\frac{z_{21}z_{23}z_{24}}{z_{31}z_{32}z_{34}}}
\ \Bigg[
\frac{\partial}{\partial z_2}
\left\langle
 \chi_0^{\pm}(z_4)\chi_0^{\pm}(z_3)\chi_0^{\pm}(z_2)\chi_0^{\pm}(z_1)
\right\rangle
+
\left\langle
 \chi_0^{\pm}(z_4)\chi_0^{\pm}(z_3)\chi_1^{\pm}(z_2)\chi_0^{\pm}(z_1)
\right\rangle
\\ \nonumber
&+&
\frac14 \left( \frac{1}{z_{21}} + \frac{1}{z_{32}} + \frac{1}{z_{24}}
\right)
\left\langle
 \chi_0^{\pm}(z_4)\chi_0^{\pm}(z_3)\chi_0^{\pm}(z_2)\chi_0^{\pm}(z_1)
\right\rangle
\Bigg].
\end{eqnarray}

\section{Superconformal blocks for R-R weights}

Any  4-point function
of R-R operators factorizes on  NS-NS states. Since in this case the supercurrent
and the superconformal modules coincide
one has
\begin{eqnarray*}
&&\hspace{-40pt}\Big\langle\chi^{+}_0(\infty)\chi^{+}_0(1)\,j_0\,\chi^{+}_0(z)\chi^{+}_0(0)\Big\rangle
\\[6pt]
&=&
\sum\limits_{p}\sum\limits_{\scriptstyle K,M,L,N}\!\!\!\!\!\!
\eta_{\infty\,1\,0} (\chi^+_0,\chi^+_0,\nu_{p,\scriptstyle KM}) B^{\scriptstyle KM,LN}
\eta_{\infty\,z\,0}(j_0\nu_{p,\scriptstyle LN},\chi^+_0,\chi^+_0),
\end{eqnarray*}
where due to the left parity conservation the sum runs over
even states, $|K|,|L|\in \mathbb{N}\cup\{0\}$. Taking into account  factorization properties
(\ref{factorization:even})
and the definitions of the superconformal blocks given in \cite{Hadasz:2006sb} one gets
\begin{eqnarray*}
&&\hspace{-40pt}\Big\langle\chi^{+}_0(\infty)\chi^{+}_0(1)\,j_0\,\chi^{+}_0(z)\chi^{+}_0(0)\Big\rangle
\\
&=&
\sum\limits_{p}
p\,\eta_{\infty\,1\,0}(\chi^+_0,\chi^+_0,\nu_{p})
\eta_{\infty\,1\,0}(\nu_{p},\chi^+_0,\chi^+_0)
\\
&&
\times\hskip -3mm \sum\limits_{\scriptstyle K,M,L,N}\!\!\!\!\!\!
\rho_{\infty\,1\,0} (\chi_0,\chi_0,\nu_{p,\scriptstyle KM}) B^{\scriptstyle KM,LN}
\rho_{\infty\,z\,0}(j_0\nu_{p,\scriptstyle LN},\chi_0,\chi_0)
\\
&=&
\sum\limits_{p}p \,C_p\,
F^{1}_{\Delta_p}\!\left[^{\Delta_0\ \Delta_0}_{\Delta_0\ \Delta_0}\right]\!(z),
\end{eqnarray*}
where $C_p \equiv \eta_{\infty\,1\,0}(\chi^+_0,\chi^+_0,\nu_{p})
\eta_{\infty\,1\,0}(\nu_{p},\chi^+_0,\chi^+_0)$. On the other hand using the
relation
$$
\eta_{\infty\,1\,0}(\nu_{p},j_{-{1\over 2}}\chi^+_0,\chi^+_0)=
p \,\eta_{\infty\,1\,0}(\nu_{p},\chi^+_0,\chi^+_0)
$$
shown in  the Appendix (\ref{C:drugi:zwiazek}) and the formula
$j_{-{1\over 2}} \chi^\pm_0 = \sqrt{2}S_{-{1\over 2}} \chi^\mp_0$
one obtains
\begin{eqnarray*}
\Big\langle\chi^{+}_0(\infty)\chi^{+}_0(1)\,j_{-{1\over 2}}\,\chi^{+}_0(z)\chi^{+}_0(0)\Big\rangle
&=&
\sum\limits_{p}p \,C_p\,
F^{1}_{\Delta_p}\!\left[^{\Delta_0 \,*\Delta_0}_{\Delta_0\ \,\Delta_0}\right]\!(z).
\end{eqnarray*}

For the sake of brevity we have ignored so far the $\bar z $ dependence of the correlation functions.
If we choose  in the right sector the fields $\chi^+_0(\bar z_i),$ the $\bar z $ dependence
of the correlator is described by
 the anti-holomorphic factor
$$
C_p\, F^{1}_{\Delta_p}\!\left[^{\Delta_0\ \Delta_0}_{\Delta_0\ \Delta_0}\right]\!(\bar z)\
$$
for each $p$.  These factors
are the same in both correlation functions appearing in
equation (\ref{first}).
Since they are linearly independent one gets from (\ref{first})
the following equation for superconformal blocks:
\begin{equation}
\label{Ia_block}
F^{1}_{\Delta_p}\!\left[^{\Delta_0\ \Delta_0}_{\Delta_0\ \Delta_0}\right]\!(z)
= {2K(z)\over \pi}\sqrt{z(1-z)}\,
F^{1}_{\Delta_p}\!\left[^{\Delta_0 \,*\Delta_0}_{\Delta_0\ \,\Delta_0}\right]\!(z) .
\end{equation}

The function $\Big\langle\chi^{+}_0(\infty)\chi^{-}_0(1)\,j_0\,\chi^{+}_0(z)\chi^{-}_0(0)\Big\rangle$
factorizes on odd states.  Using  formulae (\ref{C:pierwszy:zwiazek}) and (\ref{C:drugi:zwiazek}) one has
in this case
\begin{eqnarray*}
&&\hspace{-40pt}\Big\langle\chi^{+}_0(\infty)\chi^{-}_0(1)\,j_0\,\chi^{+}_0(z)\chi^{-}_0(0)\Big\rangle
\\
&=&
\sum\limits_{p}
p\,
\eta_{\infty\,1\,0}(\chi^+_0,S_{-{1\over 2}}\chi^-_0,\nu_{p})
\eta_{\infty\,1\,0}(\nu_{p},S_{-{1\over 2}}\chi^+_0,\chi^-+_0)
\\
&&
\times
\hskip -3mm
\sum\limits_{\scriptstyle K,M,L,N}\!\!\!\!\!\!
\rho_{\infty\,1\,0} (\chi_0,\chi_0,\nu_{p,\scriptstyle KM}) B^{\scriptstyle KM,LN}
\rho_{\infty\,z\,0}(j_0\nu_{p,\scriptstyle LN},\chi_0,\chi_0)
\\
&=&
\sum\limits_{p}p \,\Delta_p\,C_p\,
F^{1\over 2}_{\Delta_p}\!\left[^{\Delta_0\ \Delta_0}_{\Delta_0\ \Delta_0}\right]\!(z)
\end{eqnarray*}
and
\begin{eqnarray*}
\Big\langle\chi^{+}_0(\infty)\chi^{-}_0(1)\,j_{-{1\over 2}}\,\chi^{+}_0(z)\chi^{-}_0(0)\Big\rangle
&=&
\sum\limits_{p}p \,C_p\,
F^{1\over 2}_{\Delta_p}\!\left[^{\Delta_0 \,*\Delta_0}_{\Delta_0\ \,\Delta_0}\right]\!(z),
\end{eqnarray*}
which yields
\begin{equation}
\label{Ib_block}
\Delta_p\,F^{1\over 2}_{\Delta_p}\!\left[^{\Delta_0\ \Delta_0}_{\Delta_0\ \Delta_0}\right]\!(z)
\;\ = \;\
{2K(z)\over \pi}\sqrt{z(1-z)}\,
F^{1\over 2}_{\Delta_p}\!\left[^{\Delta_0 \,*\Delta_0}_{\Delta_0\ \,\Delta_0}\right]\!(z).
\end{equation}
Essentially the same method can be applied to the equation (\ref{second}). This and relations (\ref{C:drugi:zwiazek}),
(\ref{C:trzeci:zwiazek}) lead to the equations:
\newpage
\begin{eqnarray}
\label{IIa_block}
2\Delta_p\,F^{1}_{\Delta_p}\!\left[^{\Delta_0 \,*\Delta_0}_{\Delta_0\ \,\Delta_0}\right]\!(z)
&=&[z(1-z)]^{1\over 4}
{\partial\over \partial z}
\left[{2K(z)\over \pi}[z(1-z)]^{1\over 4}
F^{1}_{\Delta_p}\!\left[^{\Delta_0\ \Delta_0}_{\Delta_0\ \Delta_0}\right]\!(z)
\right]
\\
\nonumber
&+&\Delta_p
{2K(z)\over \pi} [z(1-z)]^{1\over 2}
F^{1}_{\Delta_p}\!\left[^{\Delta_0\ \Delta_1}_{\Delta_0\ \Delta_0}\right]\!(z),
\\[10pt]
\label{IIb_block}
2\,F^{1\over 2}_{\Delta_p}\!\left[^{\Delta_0 \,*\Delta_0}_{\Delta_0\ \,\Delta_0}\right]\!(z)
&=&[z(1-z)]^{1\over 4}
{\partial\over \partial z}
\left[{2K(z)\over \pi}[z(1-z)]^{1\over 4}
F^{1\over 2}_{\Delta_p}\!\left[^{\Delta_0\ \Delta_0}_{\Delta_0\ \Delta_0}\right]\!(z)
\right]
\\
\nonumber
&+&\left(\Delta_p -\textstyle{1\over 2}\right)
{2K(z)\over \pi} [z(1-z)]^{1\over 2}
F^{1\over 2}_{\Delta_p}\!\left[^{\Delta_0\ \Delta_1}_{\Delta_0\ \Delta_0}\right]\!(z).
\end{eqnarray}

The next two equations can be obtained from (\ref{third_a}) and (\ref{third_b}),
respectively:
\begin{eqnarray}
\label{IIIa_block}
&&
\hspace{-50pt}
\Delta_p\,
F^{1\over 2}_{\Delta_p}\!\left[^{\Delta_0 \ \Delta_0}_{\Delta_0\ \Delta_0}\right]\!(z)
\;=\;
(1- \sqrt{1-z})\, z^{-{1\over 2}}
F^{1}_{\Delta_p}\!\left[^{\Delta_0\ \Delta_0}_{\Delta_0\ \Delta_0}\right]\!(z)
\\[10pt]
\label{IIIb_block}
&&
\hspace{-50pt}
[z(1-z)]^{-{1\over 4}}
{\partial\over \partial z}
\left[
[z(1-z)]^{1\over 4}
F^{1}_{\Delta_p}\!\left[^{\Delta_0\ \Delta_0}_{\Delta_0\ \Delta_0}\right]\!(z)
\right]\;=
\\
\nonumber
&=&
\Delta_p(\Delta_p-{\textstyle {1\over 2}})\sqrt{z}\,
F^{1\over 2}_{\Delta_p}\!\left[^{\Delta_0 \ \Delta_1}_{\Delta_0\ \Delta_0}\right]\!(z)
+
\Delta_p
{2K(z)\over \pi} \sqrt{1-z}\,
F^{1}_{\Delta_p}\!\left[^{\Delta_0\ \Delta_1}_{\Delta_0\ \Delta_0}\right]\!(z).
\end{eqnarray}
Formulae (\ref{Ia_block}) -- (\ref{IIIa_block}) allow to express the functions
$F^{f}_{\Delta_p}\!\left[^{\Delta_0\ \Delta_1}_{\Delta_0\ \Delta_0}\right]\!(z),$
$F^{f}_{\Delta_p}\!\left[^{\Delta_0 \,*\Delta_0}_{\Delta_0\ \,\Delta_0}\right]\!(z)$
and
$F^{1\over 2}_{\Delta_p}\!\left[^{\Delta_0 \ \Delta_0}_{\Delta_0\ \Delta_0}\right]\!(z)$
in terms of
$
F^{1}_{\Delta_p}\!\left[^{\Delta_0\ \Delta_0}_{\Delta_0\ \Delta_0}\right]\!(z).
$
Using (\ref{IIIb_block}) we then arrive at the equation
\begin{eqnarray}
\label{G:equation}
\frac{d G_p(z)}{dz}
& = &
\left[
\frac{\pi^2\Delta_p}{4z(1-z)\,K^2(z)}
-\frac{1-\sqrt{1-z}}{4z\sqrt{1-z}}
\right]G_p(z),
\end{eqnarray}
where
\[
G_p(z) \;\ = \;\ [z(1-z)]^{1\over 4}\left(\frac{2K(z)}{\pi}\right)^\frac12
F^{1}_{\Delta_p}\!\left[^{\Delta_0\ \Delta_0}_{\Delta_0\ \Delta_0}\right]\!(z).
\]
Integrating (\ref{G:equation}) we get
\begin{eqnarray*}
F^{1}_{\Delta_p}\!\left[^{\Delta_0\ \Delta_0}_{\Delta_0\ \Delta_0}\right]\!(z)
& = &
(16q)^{\Delta_p}\, \left(\frac{1+\sqrt{1-z}}{2}\right)^{\frac12}\,
\left[z(1-z)\right]^{-\frac14}\,
\left(\frac{\pi}{2K(z)}\right)^{\frac12}.
\end{eqnarray*}
Using relations:
\begin{eqnarray*}
  \frac{2K(z)}{\pi} = \theta_3^2(q), \qquad
 \left(\frac{1+\sqrt{1-z}}{2}\right)^{\frac12}\, \theta_3(q) = \theta_3(q^2), \qquad
 \left(\frac{1-\sqrt{1-z}}{2}\right)^{\frac12}\, \theta_3(q) = \theta_2(q^2)
\end{eqnarray*}
where theta functions are defined in the standard way:
\begin{eqnarray*}
  \theta_3(q) = \sum_{n=-\infty}^{\infty} q^{n^2}, \qquad
 \theta_2(q) = \sum_{n=-\infty}^{\infty} q^{(n+\frac12)^2}
\end{eqnarray*}
one finally obtains:
\newpage
\begin{eqnarray}
\label{final1a}
F^{1}_{\Delta_p}\!\left[^{\Delta_0\ \Delta_0}_{\Delta_0\ \Delta_0}\right]\!(z)
& = &
 \left[z(1-z)\right]^{-\frac14}\, (16q)^{\Delta_p}\, \theta_3^{-2}(q)\, \theta_3(q^2),
\\ [6pt]
\label{final1b}
 F^{\frac12}_{\Delta_p}\!\left[^{\Delta_0\ \Delta_0}_{\Delta_0\ \Delta_0}\right]\!(z)
& = &  \left[z(1-z)\right]^{-\frac14}\, \frac{(16q)^{\Delta_p}}{\Delta_p} \, \theta_3^{-2}(q)\, \theta_2(q^2) ,
\\[6pt]
\label{final2a}
F^{1}_{\Delta_p}\!\left[^{\Delta_0 \,*\Delta_0}_{\Delta_0\ \,\Delta_0}\right]\!(z)
& = &
 \left[z(1-z)\right]^{-\frac34}\, (16q)^{\Delta_p}\, \theta_3^{-4}(q) \, \theta_3(q^2) ,
\\[6pt]
\label{final2b}
F^{\frac12}_{\Delta_p}\!\left[^{\Delta_0 \,*\Delta_0}_{\Delta_0\ \,\Delta_0}\right]\!(z)
& = &
 \left[z(1-z)\right]^{-\frac34}\, (16q)^{\Delta_p}\, \theta_3^{-4}(q) \, \theta_2(q^2),
\\[6pt]
\label{final3a}
F^{1}_{\Delta_p}\!\left[^{\Delta_0 \, \Delta_1}_{\Delta_0 \,\Delta_0}\right]\!(z)
& = &
 \left[z(1-z)\right]^{-\frac54}\, (16q)^{\Delta_p}\, \theta_3^{-6}(q) \,
\left(\theta_3(q^2) - \frac{q}{\Delta_p} \,  \frac{\partial}{\partial q} \theta_3(q^2) \right),
\\[6pt]
\label{final3b}
F^{\frac12}_{\Delta_p}\!\left[^{\Delta_0 \, \Delta_1}_{\Delta_0 \,\Delta_0}\right]\!(z)
& = &
 \left[z(1-z)\right]^{-\frac54}\, \frac{(16q)^{\Delta_p}}{\Delta_p - \frac12} \, \theta_3^{-6}(q) \,
\left(\theta_2(q^2) - \frac{q}{\Delta_p} \, \frac{\partial}{\partial q} \theta_2(q^2) \right).
\end{eqnarray}

Equations for functions $F^{f}_{\Delta_p}\!\left[^{*\Delta_0 \,*\Delta_0}_{\ \Delta_0\ \ \Delta_0}\right]\!(z)$
 can be obtained from
(\ref{czwarty}) using the relations  (\ref{C:drugi:zwiazek}), (\ref{C:trzeci:zwiazek}), (\ref{czwary:zwiazek}):
\begin{eqnarray*}
\frac{2 \Delta_p}{\sqrt{z}}\,
F^{1}_{\Delta_p}\!\left[^{*\Delta_0 \ *\Delta_0}_{\ \Delta_0\ \ \Delta_0}\right]\!(z)
& = &
\left({\partial\over \partial z} + \frac{1}{4 z(1-z)} \right)
F^{1}_{\Delta_p}\!\left[^{\Delta_0\ \Delta_0}_{\Delta_0\ \Delta_0}\right]\!(z)
+
\Delta_p \, F^{1}_{\Delta_p}\!\left[^{\Delta_0\ \Delta_1}_{\Delta_0\ \Delta_0}\right]\!(z),
\\[10pt]
 \frac{2}{\Delta_p\sqrt{z}}  \,
F^{\frac12}_{\Delta_p}\!\left[^{*\Delta_0\ *\Delta_0}_{\ \Delta_0\ \ \Delta_0}\right]\!(z)
& = &
\left({\partial\over \partial z} + \frac{1}{4 z(1-z)} \right)
F^{1\over 2}_{\Delta_p}\!\left[^{\Delta_0 \ \Delta_0}_{\Delta_0\ \Delta_0}\right]\!(z)
+
(\Delta_p - \frac12) \,
 F^{1\over 2}_{\Delta_p}\!\left[^{\Delta_0\ \Delta_1}_{\Delta_0\ \Delta_0}\right]\!(z).
\end{eqnarray*}
From the results (\ref{final1a}), (\ref{final3a}) and (\ref{final1b}), (\ref{final3b}) one gets, respectively:
\begin{eqnarray}
\nonumber
F^{1}_{\Delta_p}\!\left[^{*\Delta_0 \,*\Delta_0}_{\ \Delta_0\ \,\Delta_0}\right]\!(z)
& = & \;
z^{-\frac34}(1-z)^{-\frac54}\, (16q)^{\Delta_p} \,
\frac{\theta_3(q^2)}{\theta_3^{6}(q)}
\,
\left( 1 - \frac{q}{\Delta_p} \, \theta_3^{-1}(q) \, \frac{\partial \theta_3(q)}{\partial q}
+  \frac{\theta_2^{4}(q)}{4 \Delta_p}  \, \right),
\\[-4pt]
\label{final4}
\\[0pt]
\nonumber
F^{\frac12}_{\Delta_p}\!\left[^{*\Delta_0 \,*\Delta_0}_{\ \Delta_0\ \,\Delta_0}\right]\!(z)
& = & \!\!\!
- z^{-\frac34}(1-z)^{-\frac54}\, (16q)^{\Delta_p} \,
\frac{\theta_2(q^2)}{\theta_3^{6}(q)}\,
\Delta_p\left( 1 - \frac{q}{\Delta_p} \, \theta_3^{-1}(q) \, \frac{\partial\theta_3(q)}{\partial q}
+  \frac{\theta_2^{4}(q)}{4 \Delta_p}\, \right).
\end{eqnarray}

Explicit expressions  for the conformal blocks (\ref{final1a} -- \ref{final4}) constitute the main result
of the present work and were used in the derivation of the elliptic recurrence representation of the NS
blocks \cite{HJS}.

\section*{Acknowledgements}
The work of L.H and Z.J. was partially supported by the Polish State Research
Committee (KBN) grant no.\ 1 P03B 025 28.

\vskip 1mm
\noindent
The research of L.H.\ was supported by the Alexander von Humboldt Foundation.

\vskip 1mm
\noindent
P.S.\ is grateful to the faculty of the Institute of Theoretical Physics, University of Wroc\l{}aw,
for the hospitality.

\section*{Appendix}
\setcounter{equation}{0}
\renewcommand{\theequation}{A.\arabic{equation}}

In the Appendix we shall derive formulae expressing some three-point correlation
functions
of primary fields $j_{-\frac12}\chi_0^{\pm}(z),\ \chi_1^{\pm}(z)$ and $S_{-\frac12}\chi_1^+(z)$
through the ``basic'' three-point function
$
\big\langle
\varphi_p(z_3)\chi_0^{+}(z_2)\chi_0^{+}(z_1)
\big\rangle.
$
The used methods are simplified versions of those that led to the derivation
of equations (\ref{first}) -- (\ref{third_b}).

It follows from the OPE-s:
\begin{eqnarray*}
\psi(\xi)\chi^{\pm}(z)
& \sim &
\frac{1}{\sqrt{\xi-z}}\psi_0\chi^{\pm}(z)
\;\ = \;\
\frac{1}{\sqrt{2(\xi-z})}\chi^{\mp}(z),
\\[6pt]
\psi(\xi)\varphi_p(z)
& \sim &
1,
\end{eqnarray*}
that the function
\[
f(\xi)
\;\ = \;\
\frac{1}{\sqrt{(\xi-z_2)(\xi-z_1)}}
\Big\langle
\psi(\xi)\varphi_p(z_3)\chi_0^{-}(z_2)\chi_0^{+}(z_1)
\Big\rangle
\]
is analytic in the complex $\xi$ plane save  the simple poles at $\xi = z_2,\ \xi = z_2,$
and falls off at infinity faster than $\xi^{-1}.$ We thus have
\begin{eqnarray*}
0
& = &
\oint\limits_{z_3}\frac{d\xi}{2\pi i}\ f(\xi)
\;\ = \;\
-
\oint\limits_{z_2}\frac{d\xi}{2\pi i}\ f(\xi)
+
\oint\limits_{z_1}\frac{d\xi}{2\pi i}\ f(\xi)
\\[10pt]
& = &
-\frac{1}{\sqrt{2z_{21}}}\Big\langle
\varphi_p(z_3)\chi_0^{+}(z_2)\chi_0^{+}(z_1)
\Big\rangle
+
\frac{1}{\sqrt{2z_{12}}}\Big\langle
\varphi_p(z_3)\chi_0^{-}(z_2)\chi_0^{-}(z_1)
\Big\rangle
\end{eqnarray*}
so that
\[
\Big\langle
\varphi_p(z_3)\chi_0^{-}(z_2)\chi_0^{-}(z_1)
\Big\rangle
\;\ = \;\
\frac{\sqrt{z_{12}}}{\sqrt{z_{21}}}\,
\Big\langle
\varphi_p(z_3)\chi_0^{+}(z_2)\chi_0^{+}(z_1)
\Big\rangle
\;\ = \;\
i\Big\langle
\varphi_p(z_3)\chi_0^{+}(z_2)\chi_0^{+}(z_1)
\Big\rangle
\]
or, equivalently,
\begin{equation}
\label{C:pierwszy:zwiazek}
\eta_{z_3\,z_2\,z_1}(\nu_p,\chi_0^-,\chi_0^-)
\;\ = \;\
i\,\eta_{z_3\,z_2\,z_1}(\nu_p,\chi_0^+,\chi_0^+).
\end{equation}
Here and below we adopt the convention that for $j < l:$
\[
z_{jl} \;\ = \;\ {\rm e}^{i\pi}z_{l\!j}.
\]
Next, integrating around $\xi = z_3$ the identity
\[
\Big\langle
j(\xi)\varphi_p(z_3)\chi(z_2)\chi(z_1)
\Big\rangle
\;\ = \;\
\frac{z_{23}\sqrt{z_{21}}}{(\xi-z_3)\sqrt{(\xi-z_2)(\xi-z_1)}}
\Big\langle
\varphi_p(z_3)j_{-\frac12}\chi(z_2)\chi(z_1)
\Big\rangle
\]
we get
\[
p\,\Big\langle
\varphi_p(z_3)\chi(z_2)\chi(z_1)
\Big\rangle
\; = \;
\sqrt{\frac{z_{21}z_{32}}{z_{31}}}\,
\Big\langle
\varphi_p(z_3)j_{-\frac12}\chi(z_2)\chi(z_1)
\Big\rangle,
\]
what gives
\begin{equation}
\label{C:drugi:zwiazek}
\eta_{z_3\,z_2\,z_1}(\nu_p,j_{-\frac12}\chi_0^\pm,\chi_0^\pm)
\;\ = \;\
p\sqrt{\frac{z_{31}}{z_{21}z_{32}}}\,\eta_{z_3\,z_2\,z_1}(\nu_p,\chi_0^\pm,\chi_0^\pm).
\end{equation}
Analogous computation gives
\[
\eta_{z_4,z_3,z_2}(\chi_0^\pm,j_{-\frac12}\chi_0^\pm,\nu_p)
\;\ = \;\
-ip\sqrt{\frac{z_{42}}{z_{43}z_{32}}}\,\eta_{z_4,z_3,z_2}(\chi_0^\pm\,\chi_0^\pm,\nu_p).
\]
Using the OPE
\[
j(\xi)j_{-\frac12}\chi^{\pm}_0(z)
\;\ \sim \;\
\frac{1}{2(\xi-z)^{\frac32}}\,\chi^{\pm}_0(z)
+
\frac{1}{\sqrt{\xi-z}}\,j_{-\frac12}^2\chi^{\pm}_0(z)
\]
we next get
\begin{equation}
\label{temp:1}
\Big\langle
j(\xi) \varphi_p(z_3)j_{-\frac12}\chi^{\pm}_0(z_2)\chi^{\pm}_0(z_1)
\Big\rangle
\;\ = \;\
\left(
\frac{a}{\xi -z_2} + b
\right)
\frac{1}{(\xi-z_3)\sqrt{(\xi-z_1)(\xi-z_2)}},
\end{equation}
with
\begin{eqnarray*}
a & = &
\frac12 z_{23}\sqrt{z_{21}}
\Big\langle
\varphi_p(z_3)
\chi^{\pm}_0(z_2)\chi^{\pm}_0(z_1)
\Big\rangle,
\\[6pt]
b
& = &
z_{23}\sqrt{z_{21}}
\left[
\Big\langle
\varphi_p(z_3)j_{-\frac12}^2\chi^{\pm}_0(z_2)\chi^{\pm}_0(z_1)
\Big\rangle
+
\frac14\left(\frac{1}{z_{21}} + \frac{2}{z_{23}}\right)
\Big\langle
\varphi_p(z_3)
\chi^{\pm}_0(z_2)\chi^{\pm}_0(z_1)
\Big\rangle
\right].
\end{eqnarray*}
Integrating (\ref{temp:1}) around $\xi = z_3$ we derive a relation
\begin{equation}
\label{temp:2}
\eta_{z_3\,z_2\,z_1}(\nu_p,j^2_{-\frac12}\chi_0^\pm,\chi_0^\pm)
\;\ = \;\
\left(
\frac{2\Delta_p - \frac14}{z_{21}}
+
\frac{2\Delta_p}{z_{32}}
\right)
\eta_{z_3\,z_2\,z_1}(\nu_p,\chi_0^\pm,\chi_0^\pm).
\end{equation}
Since
\[
\eta_{z_3\,z_2\,z_1}(\nu_p,\chi_0^\pm,\chi_0^\pm)
\;\ = \;\
z_{32}^{-\Delta_p}z_{31}^{-\Delta_p}z_{21}^{\Delta_p-\frac14}\,
\eta_{\infty,1,0}(\nu_p,j_{-\frac12}\chi_0^\pm,\chi_0^\pm),
\]
we get
\begin{eqnarray}
\label{C:trzeci:zwiazek}
\nonumber
\eta_{z_3\,z_2\,z_1}(\nu_p,\chi_1^\pm,\chi_0^\pm)
& = &
\eta_{z_3\,z_2\,z_1}(\nu_p,j^2_{-\frac12}\chi_0^\pm,\chi_0^\pm)
-
\frac12
\eta_{z_3\,z_2\,z_1}\left(\nu_p,(j^2_{-\frac12}+ \psi_{-1}\psi_0)\chi_0^\pm,\chi_0^\pm\right)
\\[6pt]
& = &
\eta_{z_3\,z_2\,z_1}(\nu_p,j^2_{-\frac12}\chi_0^\pm,\chi_0^\pm)
-
\frac{\partial}{\partial z_2}
\eta_{z_3\,z_2\,z_1}\left(\nu_p,\chi_0^\pm,\chi_0^\pm\right)
\\[6pt]
\nonumber
& = &
\Delta_p
\frac{z_{31}}{z_{21}z_{32}}
\eta_{z_3\,z_2\,z_1}(\nu_p,\chi_0^\pm,\chi_0^\pm).
\end{eqnarray}
Finally, similar calculation with the help of the relation
\begin{eqnarray*}
S_{-\frac12}\chi_1^{\pm}
& = &
\left(3j_{-\frac12}^3\psi_0 + 2S_{-\frac32} -  5L_{-1}S_{-\frac12}\right)\chi_0^{\pm},
\end{eqnarray*}
gives
\begin{equation}
\label{czwary:zwiazek}
\eta_{z_3\,z_2\,z_1}(\nu_p,S_{-\frac12}\chi_1^+,\chi_0^-)
\;\ = \;
\frac{ip}{\sqrt 2}
\left(\Delta_p-\frac12\right)
\left(
\frac{z_{31}}{z_{21}z_{32}}
\right)^{\frac32}
\eta_{z_3\,z_2\,z_1}(\nu_p,\chi_0^+,\chi_0^+).
\end{equation}

\end{document}